\documentclass[%
 aip,
 jmp,%
 amsmath,amssymb,
 reprint,%
 unsortedaddress
]{revtex4-1}

\usepackage{graphicx}% Include figure files
\usepackage{dcolumn}% Align table columns on decimal point
\usepackage{bm}% bold math
\usepackage{multirow}
\usepackage{titlesec}
\titlespacing*{\subsection}
{0pt}{2.5ex}{0.5ex}
\titlespacing*{\section}
{0pt}{3.5ex}{1.0ex}

\begin{document}

%\preprint{AIP/123-QED}

\title{Measuring Multi-Configurational Character by Orbital Entanglement}

\author{Christopher J. Stein}
\author{Markus Reiher}
\email[Corresponding author: ]{markus.reiher@phys.chem.ethz.ch}
\affiliation{ 
ETH Z\"urich, Laboratorium f\"ur Physikalische Chemie, Vladimir-Prelog-Weg 2, 8093 Z\"urich, Switzerland
}

\date{\today}

\begin{abstract}
One of the most critical tasks at the very beginning of a quantum-chemical investigation is the choice of either a multi- or single-configurational method.
Naturally, many proposals exist to define a suitable diagnostic of the multi-configurational character for various types of wave functions in order to assist this crucial decision.
Here, we present a new orbital-entanglement based multi-configurational diagnostic termed $Z_{s(1)}$.
The correspondence of orbital entanglement and static (or nondynamic) electron correlation permits the definition of such a diagnostic.
We chose our diagnostic to meet important requirements such as well-defined limits for pure single-configurational and multi-configurational wave functions.
The $Z_{s(1)}$ diagnostic can be evaluated from a partially converged, but qualitatively correct, and therefore inexpensive density matrix renormalization group wave function as in our recently presented automated active orbital selection protocol.
Its robustness and the fact that it can be evaluated at low cost make this diagnostic a practical tool for routine applications.
\end{abstract}

\maketitle

\setlength{\parindent}{0cm}
\setlength{\parskip}{0.6em plus0.2em minus0.1em}

%%%%%%%%%%%%%%%%%%%%%%%%%%%%%%%%%%%%%%%%%%%%%%%%%%%%%%%%%%%%%%%%%%%%%%%%%%%%%%%
\section{Introduction}
%%%%%%%%%%%%%%%%%%%%%%%%%%%%%%%%%%%%%%%%%%%%%%%%%%%%%%%%%%%%%%%%%%%%%%%%%%%%%%%

The electronic structure of molecules is undoubtedly diverse and required the development of the plethora of quantum chemical methods applied today.
Since there is not a single (feasible) method that allows calculations of sufficient accuracy for an arbitrary problem, the choice of a suitable approach stands at the beginning of each quantum chemical investigation.
The classification according to the degree of static electron correlation of the wave function is of particular importance.
Static electron correlation occurs when the wave function must be represented by more than one electronic configuration with considerable weight, while dynamic electron correlation is caused by the multitude of configurations with little weight in the total wave function.
A robust diagnostic for the degree of static correlation is highly required.\\
Naturally, several such diagnostics were proposed to assist the selection of a suitable method for the problem under investigation.
Among these is the $T_1$-diagnostic,\cite{lee89} which is defined as the Frobenius norm of the single-excitation amplitude vector of a coupled-cluster wave function with singles and doubles excitations divided by the square root of the number of correlated electrons.
Closely related is the $D_1$ diagnostic, which is based on the matrix norm of the same single-excitation amplitude vector.\cite{jans98}
In addition to these diagnostics a density functional theory based diagnostic was proposed that quantifies the error introduced by the exact (Hartree--Fock (HF)) 
exchange in hybrid functionals, where it is known that HF exchange is inaccurate for multi-configurational systems.\cite{fogu12}

While these definitions all rely on a single-configurational wave function (that will be qualitatively wrong in the multi-configurational regime), other diagnostics are obtained from multi-configurational wave functions.
Among these are diagnostics based on natural orbital occupation numbers\cite{tish08,bao16} or the corresponding first-order reduced density matrix,\cite{loew55} and a diagnostic based on Monte Carlo configuration interaction (CI).\cite{coe14}
A comprehensive review of these diagnostics including a comparison for several critical cases can be found in Ref. \citenum{coe15}.
Recently, a diagnostic for static correlation from finite-temperature density functional theory (DFT) 
was presented\cite{grim15} with which the spatial location of the statically correlated electrons can be determined.

Another way to think about electron correlation is orbital entanglement derived from grand-canonical reduced density matrices.\cite{lege03,lege06,riss06}
Orbital entanglement is directly related to static (or nondynamic) correlation\cite{bogu12} and permits us here to present a multi-configurational diagnostic based on these quantities. 
Although similar attempts have already been made,\cite{lege04,bogu12,bogu14,coe15} we show how a modified diagnostic based on the orbital entanglement overcomes the problems of these previous definitions.
We furthermore show how the evaluation of this diagnostic can be incorporated in the algorithm of our recently proposed automated selection of active orbitals \cite{stei16} and can therefore be obtained at low cost.

Certainly, a strong multi-configurational character of a wave function does not imply that only multi-configurational methods yield qualitatively correct results.
If, for some specific molecule, HF orbitals are a proper one-particle basis and a high-order excitation operator is applied (say, quadruples), the single-reference
coupled cluster approach may be reliable. Also unrestricted (broken-symmetry) HF and Kohn--Sham DFT can capture strong static correlation (at the cost
of a symmetry violation) \cite{nood79,nood81,nood82,jaco12,tsuc09}.
Occasionally, however, a single-configurational method such as coupled cluster with singles, doubles and perturbative triples excitations may fail even for low values of, e.g., the $T_1$ diagnostic (cf., the F$_2$ molecule\cite{kart06}).
The authors of Ref. \citenum{kart06} therefore suggested an energy-based diagnostic that indicates when higher-order terms in the coupled-cluster expansion are required to achieve a certain accuracy.

The purpose of the diagnostic presented here is to measure the degree of static (or nondynamic) correlation.
It is, however, outside the scope of our diagnostic to suggest an optimal method only by means of certain thresholds and for all kinds of properties.
We require our measure to be capable of identifying those cases that are a) mainly dynamically correlated or b) strongly statically correlated.
Our measure should certainly suggest a careful inspection for cases that cannot be classified as one of these two limiting cases.

This article is organized as follows:
We first summarize properties we demand of a multi-configurational diagnostic and show how these requirements are met by an entanglement based criterion.
Then, we show how this diagnostic can be easily obtained in the framework of our automated orbital selection protocol and study several critical cases.

%%%%%%%%%%%%%%%%%%%%%%%%%%%%%%%%%%%%%%%%%%%%%%%%%%%%%%%%%%%%%%%%%%%%%%%%%%%%%%%
\section{Entanglement Based Multi-Configurational Diagnostic}
%%%%%%%%%%%%%%%%%%%%%%%%%%%%%%%%%%%%%%%%%%%%%%%%%%%%%%%%%%%%%%%%%%%%%%%%%%%%%%%

\subsection{Desirable properties}
By definition, a "diagnostic" has to reveal the nature of a given problem (or an aspect thereof) such that suitable measures can be taken to solve the problem.
Therefore, the main requirement for a multi-configurational diagnostic is the clear identification of strong static correlation that requires a multi-configurational description.\\
In this sense, well-defined limits are preferable and we require our measure to give a value of zero in the absence of electron correlation (i.e., when the wave function is exactly described by a single Slater determinant) and a value of one for strong static correlation.
These well-defined limits facilitate the definition of a threshold value for the diagnostic below which single-configurational methods can safely be used whereas multi-configurational methods are necessary if the diagnostic gives a value above that threshold.\\
If the diagnostic is meant to guide the selection of a suitable method rather than only assess the quality of an already converged calculation, its evaluation should take only a small fraction of the total computational time.
That a multi-configurational diagnostic should be obtained from a qualitatively correct multi-configurational wave function (that still includes single-configurational wave functions as limiting cases\cite{stei16a}) is an additional natural criterion.

\subsection{Definition and constraints}
%irgendwo muss stehen (am besten im Absatz vorher), dass das Entanglement proportional zur statischen Korrelation ist
Our multi-configurational diagnostic is based on the single-orbital entropy defined as\cite{lege03,lege06,riss06}
\begin{equation}
s_i(1) = - \sum_{\alpha = 1}^4 \omega_{\alpha,i} \ln \omega_{\alpha,i},
\end{equation}
where the $\omega_{\alpha,i}$ are the eigenvalues of the one-orbital reduced density matrix for the $i$th orbital and $\alpha$ runs over the four possible occupations in a spatial orbital basis (doubly occupied, spin up, spin down, unoccupied).
Maximum entanglement corresponds to a situation where all occupations are equally likely ($\omega_{\alpha, i} = 0.25$ for all $\alpha$) and is therefore equal to $\ln 4 \approx 1.39$.
This theoretical maximum allows for a scaling of the multi-configurational diagnostic $Z_{s(1)}$ such that correct limits as discussed above  (0 = no entanglement, 1 = full entanglement) exist,
\begin{equation}
Z_{s(1)} = \frac{1}{L \ln 4}\sum_i^L s_i(1),
\label{Z}
\end{equation}
where $L$ is the number of orbitals considered for the evaluation of $Z_{s(1)}$.
This scaling allows us to compare the diagnostic between different active space sizes which was not the case for some previous definitions of entanglement based diagnostics.\cite{lege04,bogu14} 
The single-orbital entropies can easily be evaluated from a density matrix renormalization group (DMRG)\cite{whit92,whit93,lege08,chan08,chan09,reih10,chan11,mart11,scho11,kura14,wout14,yana15,szal15,knec16,chan16} wave function but can also be implemented for other methods like standard complete active space 
self-consistent field (CASSCF) calculations or the antisymmetric product of the 1-reference orbital-based geminals\cite{bogu16}.

Maximum entanglement can only be realized if the number of electrons equals the number of spatial orbitals over which they are distributed and the number of orbitals is even.
It is therefore necessary to restrict the set of orbitals whose single-orbital entropies define $Z_{s(1)}$ to the set of most entangled orbitals of a given calculation.
If no such restriction is applied (as in Ref. \citenum{coe15}), the large number of virtual orbitals will artificially lower $Z_{s(1)}$ because the four possible occupations cannot be realized with the electrons available in the system.\\
The identification of the most entangled orbitals of a given calculation is therefore crucial to define $Z_{s(1)}$ such that its values can be compared between different systems or orbital bases.
This is also central to our recently proposed protocol for the automated selection of active orbital spaces for multi-configurational calculations.\cite{stei16}
For the calculation of $Z_{s(1)}$ we may exploit the protocol of Ref. \citenum{stei16} for partially converged, qualitatively correct DMRG wave functions to identify a set strongly entangled orbitals.
We then restrict the set of orbitals to fulfill the requirement that the number of electrons equals the number of orbitals and further exclude singly-occupied molecular orbitals (SOMOs) in open-shell cases.
This special treatment of SOMOs is also a peculiarity of other multi-configurational diagnostics and justified by the observation that they are usually only weakly entangled.\cite{tish08}\\
The diagnostic $Z_{s(1)}$ is then evaluated from this restricted set of orbitals.
Within our automated selection protocol, we obtain the diagnostic for free as a byproduct of an initial, partially converged DMRG calculation with a large active space.

%%%%%%%%%%%%%%%%%%%%%%%%%%%%%%%%%%%%%%%%%%%%%%%%%%%%%%%%%%%%%%%%%%%%%%%%%%%%%%%
\section{Computational methodology}
%%%%%%%%%%%%%%%%%%%%%%%%%%%%%%%%%%%%%%%%%%%%%%%%%%%%%%%%%%%%%%%%%%%%%%%%%%%%%%%

Computational details of calculations from previous work are described where needed.
We applied the following computational methodology for all new calculations. 
All structures were optimized with DFT in \textsc{Turbomole v6.5}\cite{ahlr89} with the PBE functional\cite{perd96,perd97} and the def2-TZVP basis set\cite{weig05}.
We applied the ANO-RCC basis set \cite{roos04,widm90} in its double zeta contraction in combination with the Douglas-Kroll-Hess Hamiltonian at second order \cite{hess86,reih04,reih04a} in all wave function calculations.
Initial orbitals were obtained with CASSCF as implemented in \textsc{Molcas 8}.\cite{aqui8}
Split-localized orbitals\cite{oliv15} were obtained after Pipek and Mezey.\cite{pipe89}
All DMRG calculations as well as the evaluation of $s(1)$ were performed with our DMRG program QCMaquis.\cite{dolf14,kell15,kell16}
For these calculations, we adopt the notation of Ref.\citenum{stei16}: DMRG[$m$]($N,L$)\#\textit{orbital\_basis}, where $m$ is the number of renormalized states, $N$ and $L$ are the number of active electrons and orbitals, respectively, and the string after the hashtag specifies the orbital basis of the DMRG-CI calculations.
With CAS($N,L$)SCF, we adopt a very similar notation to specify the setup of the initial orbital generation.
The number of sweeps was set to twelve in all DMRG calculations and the definition of a plateau in the threshold diagrams of our automated orbital selection procedure\cite{stei16} is set to ten.
When $Z_{s(1)}$ is evaluated for wave functions from previously published studies, we refer to the original literature for the computational setup 
and provide only a minimal description.

\begin{figure*}[htb]
\includegraphics[width=\textwidth]{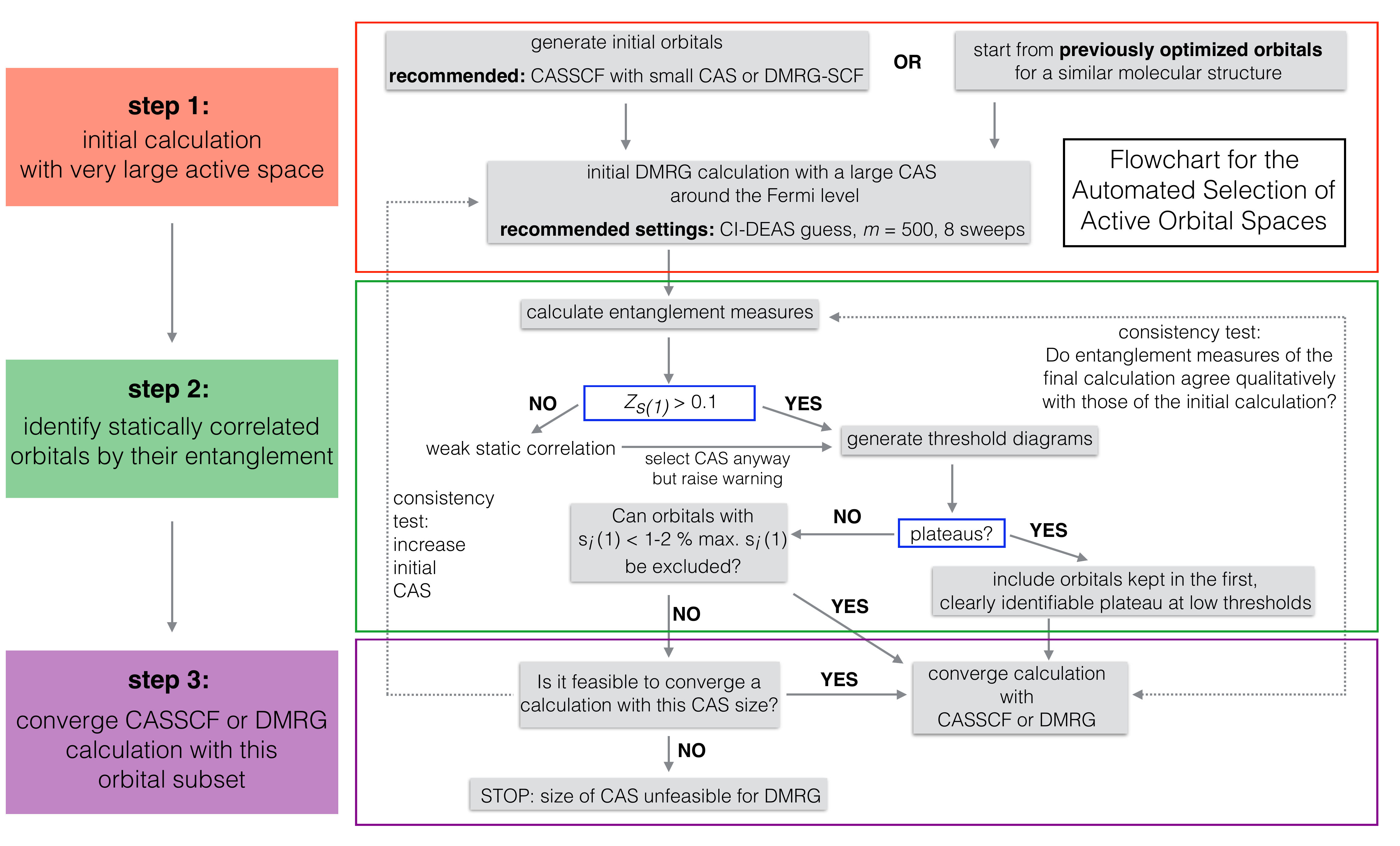}
\caption{Flowchart for the procedure of the automated active orbital space selection.}
\label{flowchart}
\end{figure*}

The automated active orbital selection exploits the benefit of DMRG to handle large active spaces with up to 100 orbitals and the fact that convergence of entanglement measures on which the method relies is faster than convergence of the energy.
The protocol can be divided into three steps (see Fig. \ref{flowchart}):
At first a partially converged but qualitatively correct DMRG wave function is calculated with a large active space including orbitals that are likely to be statically correlated.
These orbitals can be the full-valence space or in the case of transition-metal complexes the metal-centered orbitals and those valence orbitals located at the first ligand shell.
In a second step, the single-orbital entropy of the initial orbitals is calculated from the DMRG wave function.
The identification of the strongly statically correlated orbitals involves the analysis of so-called threshold diagrams\cite{stei16} in which the single-orbital entropies of the orbitals are ranked against the highest single-orbital entropy present in the wave function.
The automated protocol issues a message when the static correlation is low and single-configurational methods might be more efficient, because dynamical correlation is much more efficiently captured in, e.g., single-reference coupled-cluster calculations than for example in multi-configurational perturbation theory.
Since $Z_{s(1)}$ is calculated from the single-orbital entropies only, it can be evaluated without any additional cost in the automated selection procedure and we base the low static correlation message now on this diagnostic.
It might also be of value to adapt the length of a plateau in the threshold diagrams that defines a distinct subset of strongly statically correlated orbitals to $Z_{s(1)}$ in order to adapt the procedure even more to the degree of static correlation present in the molecule.
Although we showed in a recent study\cite{stei16a} that such an adaptation can be very beneficial, this needs to be investigated further in future work.
The last step of the automated protocol is the fully converged final calculation with the set of highly entangled orbitals only.
Consistency tests that compare the entropy information of the initial and the fully converged wave function ensure that artefacts from unbalanced active spaces are avoided and all highly entangled orbitals are still included in the final calculation.

\section{Results}
\subsection{Bond stretching}

\begin{figure}[h!]
\centering
\includegraphics[width=0.5\textwidth]{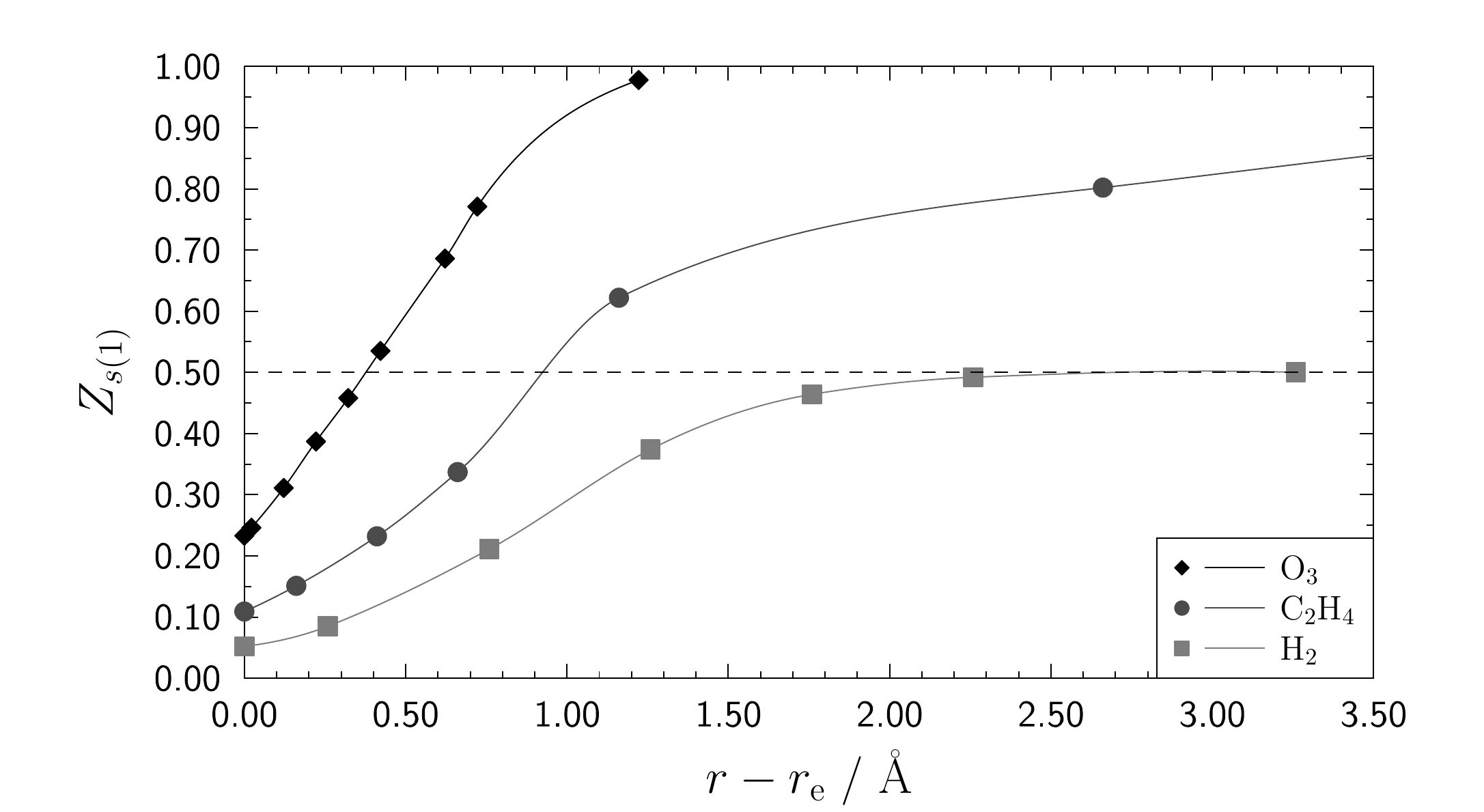}
\caption{
Multi-configurational diagnostic $Z_{s(1)}$ as a function of the HH, CC, and symmetric OO bond elongation in H$_2$, ethylene, and ozone, respectively.
The equilibrium distances are chosen as in Ref. \citenum{tish08}: $r_\mathrm{e} \mathrm{(HH)} = 0.741$~\AA,  $r_\mathrm{e} \mathrm{(CC)} = 1.339$~\AA, and $r_\mathrm{e} \mathrm{(CC)} = 1.278$~\AA.
Internal coordinates that were kept fixed are: $r_\mathrm{e} \mathrm{(CH)} = 1.086$~\AA, $\angle$HCH = 117.6$^\circ$, $\angle$HCC = 121.2$^\circ$, and $\angle$OOO = 116.8$^\circ$.
The lines are produced from spline fits and meant to guide the eye.
The number of orbitals whose entropy is included in the calculation of $Z_{s(1)}$ is two, four, and six for H$_2$, C$_2$H$_4$, and O$_3$, respectively.}
\label{limits}
\end{figure}

In many cases the limit of almost pure dynamical correlation and strong static correlation can be realized by different structures of the same molecule.
A prototype is the H$_2$ molecule, which has almost no static correlation in its equilibrium structure and becomes more and more statically correlated when the HH bond is elongated.
A similar behavior is observed for the CC stretch coordinate in ethylene and the symmetric OO stretch in ozone although the degree of static correlation is higher already in the equilibrium structure of these molecules.
Ref.\,\citenum{tish08} contains a diagram of the dependence of an occupation number based multi-configurational diagnostic on these stretch coordinates that we reproduce in Fig.\,\ref{limits} for $Z_{s(1)}$.

The figure shows that $Z_{s(1)}$ is a continuous function with increasing distances because the number of orbitals included in the calculation of $Z_{s(1)}$ does not change upon bond stretching in all cases (see caption of Fig.\ \ref{limits}) 
and gives the right order of multi-configurational character at the equilibrium structure of these three molecules.
Note that the orbitals $L$ included in the evaluation of $Z_{s(1)}$ do not include the whole set of initial CASSCF orbitals but only the subset of highly entangled orbitals identified by the automated active space selection procedure with the additional constraint that the number of orbitals equals the number of electrons.
The data in Fig.\ \ref{limits} are obtained from full-valence DMRG[500]-CI calculations with CAS(2,2)-SCF, CAS(4,4)-SCF, and CAS(6,9)-SCF initial orbitals for H$_2$, ethylene, and ozone, respectively\cite{tish08}.
Three configuration state functions (CSFs) can be generated from the full valence CAS(2,2) for H$_2$ in a singlet state with orbitals of $\sigma_g$ and $\sigma_u$ symmetry.
In these CSFs, the orbitals are either doubly occupied or singly occupied with different spin, where the latter leads to a different total symmetry so that it is effectively not realized and the wave function can be interpreted as an effective seniority-zero wave function\cite{byta11}. 
The orbitals are therefore only either doubly occupied or empty, so that the limit for $s(1)$ becomes $\ln 2$ 
which explains the maximum value of $Z_{s(1)} = 0.5$ for H$_2$ in Fig.\ \ref{limits}.
This lower limit occurs whenever symmetry constraints for the highly entangled orbitals from which $Z_{s(1)}$ is calculated restrict the CI space to seniority-zero determinants only (see the example of singlet dioxygen in Table\ \ref{s-t}).
Although the normalization in Eq.~(\ref{Z}) could be changed to $\ln 2$ for seniority-zero wave functions, we will refrain from this alternative definition in the context of our automated selection protocol because this situation occurs only for highly symmetric (and hence for mostly small) molecules.

\subsection{Size independence}
Independence of the size of the molecule or the active space is another criterion to be fulfilled by a multi-configurational diagnostic.
In order to show that this holds true for $Z_{s(1)}$, we evaluated its value for alkane chains with one, three, five, seven, and nine carbon atoms from full-valence DMRG[500]\#CAS(4,4)SCF calculations.
The initial four active orbitals were selected from around the Fermi level (i.e. HOMO-1, HOMO, LUMO, LUMO+1) because we observed in a previous study\cite{stei16a} that CASSCF orbitals selected in this way can be a suitable basis for the automated active space selection procedure. 
Certainly, this choice cannot be generalized as it might perform poorly for molecules with complicated electronic structures or excited states.
However, in the case of the alkanes the resulting orbitals are a suitable basis for the automated active orbital selection.

\begin{table}[h!]
\begin{center}
\caption{Multi-configurational diagnostic $Z_{s(1)}$ for five alkane molecules from full-valence DMRG[500]($N,L$)\#CAS(4,4)SCF calculations and four (poly)acenes from full-valence DMRG[500]($N,L$)\#CAS($N$,$L$)SCF calculations where the active orbitals consist of the $p_z$-orbitals that form the aromatic system. The number of orbitals whose entropies are included in the calculation of $Z_{s(1)}$ equals $L$.}
\begin{tabular}{cccccc}
\hline \hline
                                 & methane &propane & pentane & heptane & nonane\\
 \hline
 ($N,L$)           & (8,8)  &(20,20) & (32,32) & (44,44) &(56,56) \\
 $Z_{s(1)}$               & 0.04 & 0.03  & 0.03 & 0.03 &0.03  \\
 \hline
  & benzene & naphtalene & anthracene &tetracene & pentacene\\
($N,L$)        & (6,6)  &(10,10) & (14,14) & (18,18) & (22,22) \\
 $Z_{s(1)}$               & 0.21 & 0.23  & 0.24 & 0.25 & 0.24  \\
 
 \hline \hline
 \end{tabular}
  \label{independence}
 \end{center}
\end{table}

In all cases, the diagnostic has a very low value and is almost exactly the same for these three molecules with very similar electronic structure.
We obtain identical results for all five alkanes although only the calculation on methane is converged with respect to the energy with 500 renormalized states of the DMRG calculation.
The active space selected by the automated procedure is the full-valence space.
This is due to the fact that even the most entangled orbitals have a very low single-orbital entropy and a selection with respect to the maximum $s(1)$ value of a given calculation is prone to overestimate the number of active orbitals required.
In these cases, a very low $Z_{s(1)}$ value can in the automated selection procedure\cite{stei16} point toward  single-configurational methods, that are then a more appropriate choice.
We further analyzed the multi-configurational character of the electronic ground state of several polyacenes in Table \ref{independence}.
In contrast to the series of alkane molecules, we observe a progressively strong multi-configurational character in agreement with previous studies\cite{bend04,hach07,gido08,hajg09,pelz11,mizu13,horn14,horn15,foss16}. 
However, we see a saturation because our diagnostic condenses information from all orbitals into one number while previous studies report natural orbital occupation numbers (NOONs) of all $\pi$-orbitals.
Although the multi-radical character is increased upon enlargement of the $\pi$-system, this affects not all orbitals uniformly and the effect of the additional orbitals with little entanglement counterbalances the effect of the highly entangled orbitals with radical character.
Hence, the information from the NOONs and $Z_{s(1)}$ is complementary for these molecules.

\subsection{Singlet vs. triplet wave functions}
Often, the multi-configurational character of a wave function changes significantly with the spin state.
A well-known example is the oxygen molecule that has a strong multi-configurational character in its singlet state, while a single Slater determinant is a qualitatively correct approximation to its triplet ground state.
In CF$_2$, this change of the multi-configurational character is much more subtle and in H$_2$CC, it is also less pronounced but the tendency is reversed.
The data in Table \ref{s-t} reveal that $Z_{s(1)}$ identifies very subtle differences in the multi-configurational nature of the wave function.

\begin{table}[htb]
\caption{Multi-configurational diagnostic $Z_{s(1)}$ for the singlet and triplet wave function of three molecules along with the weights of the largest CI coefficients from initial CASSCF calculations.
Structures and active spaces for the initial CASSCF calculations were chosen as in Ref. \citenum{bao16}. 
The number of orbitals whose entropies are included in the calculation of $Z_{s(1)}$ are given in parentheses.}
\begin{tabular}{cccccc}
\hline \hline
&  &\multicolumn{2}{c}{singlet} & \multicolumn{2}{c}{triplet} \\
 molecule &CAS($N,L$) & $Z_{s(1)}$  & weight  &  $Z_{s(1)}$   & weight  \\
 \hline
 O$_2$     &(12,14) & 0.37$^a$ \textbf{(2)} & 45 \%, 45 \% & 0.09\textbf{(10)}  & 91 \% \\
 CF$_2$   &(18,12) & 0.10 \textbf{(6)} & 95 \% & 0.09 \textbf{(4)} & 96 \% \\
 H$_2$CC& (6,8) & 0.10 \textbf{(8)} & 93 \% & 0.16 \textbf{(4)} & 92 \% \\
 \hline \hline
 \multicolumn{6}{l}{$^a$The limit of $Z_{s(1)}$ is 0.5 here because symmetry}\\
  \multicolumn{6}{l}{constraints lead to an effective seniority-zero}\\
   \multicolumn{6}{l}{wave function.}
 \end{tabular}
\label{s-t}
\end{table}

The information obtained agrees with that from the weights of the largest CI coefficients in the preceding CASSCF calculation, where weights that differ substantially from unity indicate multi-configurational character.\cite{lee89}
The information of the weight of the largest CI coefficient is not always available or reliable because the orbital basis is not necessarily obtained with CASSCF or with a too small active space.
Although the CI weights can also be evaluated from a DMRG wave function\cite{mori07,bogu11}, for large active spaces this procedure involves a sampling over Slater determinants with correct symmetry within the active space which is an additional (and potentially time-consuming) step.
On the contrary, the information about the multi-configurational character is more compressed in our orbital-based diagnostic and can further be evaluated without additional calculations in our automated active space selection protocol.

\subsection{Dependence on the orbital basis}
In Ref. \citenum{stei16} we investigated the suitability of different orbital bases for the automated active space selection.
Here, we examine how much the orbital basis influences the final value of $Z_{s(1)}$.
We chose MnO$_4^-$ with HF, split-localized, CAS(10,10)SCF, and DMRG[500](38,25)-SCF orbital bases as an example.
For this molecule, we evaluate $Z_{s(1)}$ from the orbital bases as described in Ref. \citenum{stei16}.
In that paper, the structure was optimized with DFT in \textsc{Turbomole v.6.5}\cite{ahlr89} with the B3LYP functional\cite{lee88,step94} and the def2-TZVP basis set\cite{weig05}.
The orbitals were obtained as described in the computational methodology section of the present article but with the ANO-S\cite{pier95} atomic orbital basis set.
We further calculated $Z_{s(1)}$ for benzene and linear hydrogen chains with eight and 16 hydrogen atoms and different bond distances (1.5~\AA \, and 2.0~\AA) as prototypes for strongly correlated model systems\cite{hach06,tsuc09,sini10,lin11,stel11} 
all with the same types of orbital bases. 
We observe only a minor basis set dependence of $Z_{s(1)}$ as it varies at most by 0.12.
Within this series of orbital bases, $Z_{s(1)}$ is largest for the CASSCF and DMRG-SCF bases (see Table~\ref{orbital_bases}).
Since these orbital bases are the natural bases of the underlying electronic structure method we note that bases that deviate from the natural basis tend to slightly underestimate the multi-configurational character of a wave function as measured by $Z_{s(1)}$.

\begin{table}[htb]
\begin{center}
\caption{Multi-configurational diagnostic $Z_{s(1)}$ for several molecules obtained from full-valence DMRG-CI calculations employing different orbital bases. The number of orbitals whose entropy is included in the calculation of $Z_{s(1)}$ is ten, six, eight, and sixteen for MnO$_4^-$,  C$_6$H$_6$, H$_8$, and H$_{16}$, respectively.}
\begin{tabular}{lcccc}
\hline \hline
         & split- &HF & CASSCF &DMRG[500]- \\
          & localized & &  & SCF\\
          \hline
\multirow{2}{*}{MnO$_4^-$}     &       &           & CAS(10,10) & CAS(38,25)\\               
                       & 0.31 & 0.33 & 0.35 & 0.38 \\
\multirow{2}{*}{C$_6$H$_6$} &  &  &CAS(6,6)  &CAS(30,30) \\
                      &  0.17 & 0.17 & 0.20     & 0.19 \\
\multirow{2}{*}{H$_8$ (1.5~\AA)} &  & & CAS(8,8) & \\
                            & 0.33 & 0.34  & 0.43 & - \\
H$_8$ (2.0~\AA) & 0.64 & 0.69 & 0.75 & -\\
\multirow{2}{*}{H$_{16}$ (1.5~\AA)} &  & & CAS(16,16) & \\
                            & 0.32 & 0.35  & 0.42 & - \\
H$_{16}$ (2.0~\AA) & 0.49 & 0.59 & 0.61 & -\\
 \hline \hline
 \end{tabular}
\label{orbital_bases}
\end{center}
\end{table}

%%%%%%%%%%%%%%%%%%%%%%%%%%%%%%%%%%%%%%%%%%%%%%%%%%%%%%%%%%%%%%%%%%%%%%%%%%%%%%%
\section{Conclusions}
%%%%%%%%%%%%%%%%%%%%%%%%%%%%%%%%%%%%%%%%%%%%%%%%%%%%%%%%%%%%%%%%%%%%%%%%%%%%%%%
We presented a new orbital entanglement based multi-configurational diagnostic and applied it to several critical examples.
Although a similar diagnostic has previously been proposed,\cite{coe15} our definition overcomes the lack of well-defined limits.
It satisfies all criteria a multi-configurational diagnostic should meet, mostly by construction, as shown in the numerical examples.
Furthermore, it can be evaluated as a by-product of our automated orbital selection protocol.\cite{stei16}
There, it even serves for the assessment of the applicability of a multi-configurational method and directly guides the choice of a computational method.
The fact that this diagnostic can be evaluated from partially converged and therefore inexpensive calculations makes it attractive for routine calculations.

A multi-configurational diagnostic is not only supposed to rank wave functions according to their multi-configurational character but should also give advice if a single- or multi-configurational method is appropriate for a given calculation.
This means that threshold values need to be introduced that split the range of values that $Z_{s(1)}$ can take into regimes that can safely be treated by a single-configurational method and those where multi-configurational methods are required.

Until today no fundamental threshold has been defined and all values suggested in the literature rely on experience.
In order to challenge our multi-configurational diagnostic, we carried out new calculations and re-evaluated recent work.
Based on these results, single-configurational methods will be reliable, if $Z_{s(1)}$ is between 0 and 0.1, whereas we advice to apply multi-configurational methods whenever this value lies between 0.2 and 1.0.
It is important to emphasize, however, that $Z_{s(1)}$ is a diagnostic for the degree of static correlation.
It does not imply that single-reference methods will fail unavoidably when the $Z_{s(1)}$ diagnostic reaches a certain value.
The ability of a single-reference method to describe wave functions with a $Z_{s(1)}$ value higher than 0.1 will, in general, be strongly depending
on the particular method chosen (e.g., on whether HF orbitals from a suitable one-electron basis or on whether a high excitation operator was chosen), 
whereas multi-reference methods allow for a generally rigorous treatment of static correlation.
The trade-off that dynamical correlation is usually less efficiently calculated than in single-reference methods has to be kept in mind and demands a careful selection of the method especially for wave functions with intermediate static correlation in the range $0.1 \leq Z_{s(1)} < 0.2$.
Our automated selection protocol\cite{stei16} will issue an automated message if $Z_{s(1)}$ should fall in that regime.
Knowing that for certain molecular structures single-reference methods can be applied does not imply that one may eventually benefit from the 
more efficient capture of dynamic correlation by single-reference coupled cluster theory as a mixture of methods will produce kinks in the potential energy surface.

%%%%%%%%%%%%%%%%%%%%%%%%%%%%%%%%%%%%%%%%%%%%%%%%%%%%%%%%%%%%%%%%%%%%%%%%%%%%%%%
\section*{Acknowledgments}
%%%%%%%%%%%%%%%%%%%%%%%%%%%%%%%%%%%%%%%%%%%%%%%%%%%%%%%%%%%%%%%%%%%%%%%%%%%%%%%
This work was supported by the Schweizerischer Nationalfonds (Project No. 200020\_169120).
C.J.S. gratefully acknowledges a K\'ekule fellowship from the Fonds der Chemischen Industrie.

%%%%%%%%%%%%%%%%%%%%%%%%%%%%%%%%%%%%%%%%%%%%%%%%%%%%%%%%%%%%%%%%%%%%%%%%%%%%%%%
\section*{References}
%%%%%%%%%%%%%%%%%%%%%%%%%%%%%%%%%%%%%%%%%%%%%%%%%%%%%%%%%%%%%%%%%%%%%%%%%%%%%%%

%\bibliography{references} %your .bib file
%merlin.mbs aipnum4-1.bst 2010-07-25 4.21a (PWD, AO, DPC) hacked
%Control: key (0)
%Control: author (8) initials jnrlst
%Control: editor formatted (1) identically to author
%Control: production of article title (0) allowed
%Control: page (1) range
%Control: year (1) truncated
%Control: production of eprint (0) enabled
%

\end{document}